\documentclass[%
 reprint,
 amsmath,amssymb,aps,
 prl,twocolumn,superscriptaddress
]{revtex4-1}
\usepackage{graphicx}
\usepackage{xcolor}
\usepackage[utf8]{inputenc}
\usepackage[T1]{fontenc}
\usepackage{amsthm}
\usepackage{ulem}
\usepackage{float}

\begin{document}

\title{Hypersonic heat-induced flows of magnons induced by femtosecond laser pulses}

\author{Sergiu Ruta}
\email{sergiu.ruta@york.ac.uk}
\affiliation{Department of Physics, University of York, York, United Kingdom}
\author{Zuwei Fu}%
\affiliation{Tongji University, Shanghai, China}
\affiliation{Department of Physics, University of York, York, United Kingdom}
\author{Thomas Ostler}%
\affiliation{Sheffield Hallam University, Sheffield, United Kingdom}
\author{Alexey Kimel}%
\affiliation{Institute for Molecules and Materials, Radboud University Nijmegen, Nijmegen,Netherlands}
\author{Roy Chantrell}%
\affiliation{Department of Physics, University of York, York, United Kingdom}


\begin{abstract}
In this work, we present evidence for the existence of a magnonic current on the sub-picosecond time-scale in a ferrimagnetic bilayer and its effect on ultrafast spin dynamics.
The ferrimagnet, GdFeCo, is a material known to undergo ultrafast switching within 1-2ps after excitation with femtosecond laser pulses. Here, we show that the strong thermal gradients induced by applying femtosecond laser pulses and the presence of chemical inhomogeneities lead to local imbalances in the effective temperatures of the spins that produces a rapid transfer of spin angular momentum, which we interpret as an ultrafast spin Seebeck effect.
We have quantified the typical magnon propagation in such a system. The results show ballistic magnon propagation with 30nm/ps velocities. The characteristic time scale of such magnon propagation indicates that this magnon transport can play an important role in switching, a crucial piece of understanding towards realising next generation data processing devices that operate at much higher frequencies.
\end{abstract}

\flushbottom
\maketitle

\thispagestyle{empty}

Transferring information using coherent high-frequency oscillations in magnetic systems (magnons) is the holy grail of low energy data processing. However, high frequency generally means short life-times of excitations making them difficult to utilise. Understanding the fundamental physics of energy transfer in spin textures via magnon flows and domain wall motion are critical for information processing~\cite{Bauer2012,Azim2014} in fields such as magnonics~\cite{Chumak2015}  and spintronics~\cite{Schellekens2014, el2019ultrafast, wang2020magneto, grundler2016spintronics}.
Because magnonic currents rely on localised magnetic moments, the motion of itinerant electrons, which come with large amounts of Joule heating losses, is not required and, importantly, the materials are not restricted to metallic system. 
This allows the use of insulating materials such as yttrium–iron–garnet (YIG), which, due to extremely low magnetic damping, has been at the core of spintronic devices. 
Furthermore, it is also possible to use magnonic currents to move topological spin textures~\cite{jiang2013direct}, such as domain walls, which have been proposed as the key component of devices such as 3D racetrack memory~\cite{parkin2008magnetic,parkin2015memory} and magnetic-based logic gates~\cite{allwood2005magnetic}. Driving towards high frequency domain wall motion via magnon currents opens up the possibility of novel memory-in-logic and logic-in-memory devices \cite{Chumak2015}, highlighting the importance of high velocity magnon excitation and manipulation on the nanometre scale.

In recent years, femtosecond lasers have been proposed as effective stimuli  for inducing fast domain wall motion, as well as, spin and magnonic current generation. Since the pioneering work of Beaurepaire et.al \cite{Beaurepaire1996}, and then the discovery of single pulse all optical switching (AOS) in GdFeCo \cite{Ostler2012c,Radu2011}, these laser pulses have been considered as the most likely candidates to achieve high-speed \cite{atxitia2018ultrafast,khorsand2012role,kimel2005ultrafast,Stanciu2007}, energy-efficient \cite{yang2017ultrafast}, field-free magnetisation switching mechanism \cite{radu2015ultrafast,Moreno2017,mangin2014engineered,kirilyuk2010ultrafast,nieves2016modeling,Koopmans2010,Bossini2018}. In a recent review, Kimel and Li~\cite{kimel2019writing} have concluded that the AOS mechanism outperforms other methods in terms of the speed of the  recording event and its unprecedentedly low heat load. 
In ultrafast magnetism there is experimental evidence for the existence of non-local angular momentum transfer\cite{Graves2013}, which was ascribed to the generation of spin currents \cite{battiato2012theory}.
However, contrary to magnon currents, 
spin current does not necessary transfer energy. Therefore it is not clear to what extent these currents are of any relevance for ultrafast information processing and transfer.

In this Letter, we show that ultrafast magnetization reversal is accompanied by magnon currents having a non-zero Poynting vector. The magnons in the flow  propagate ballistically, having a huge speed of 30 km/s i.e. must be faster than the fastest domain
wall ever observed in GdFeCo ~\cite{kim2020fast}.
We demonstrate that the magnon current has a critical role in deterministic switching of the bilayer systems and ferrimagnetic alloys, but on a localised level. The characteristic length-scale of the excitations is of the order of a few nanometres, which is
extremely promising for utilising this mechanism in magnetic element devices at the nanometre scale.

\begin{figure*}
\centering
     \includegraphics[angle = -0,width = 1\textwidth]{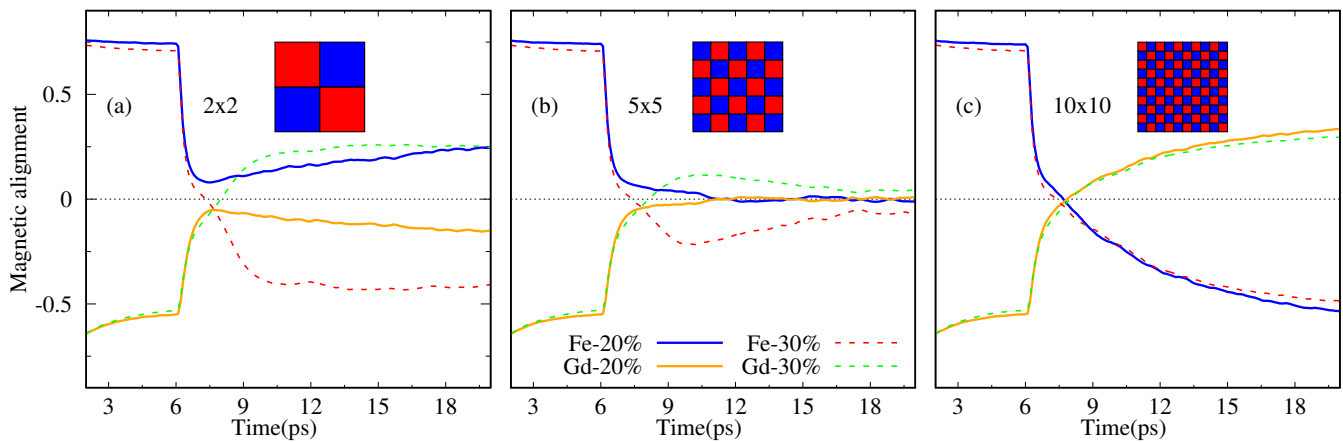}
  \caption[]{
 Checkerboard system with (a) 2X2 (b) 3X3 and (c) 10x10 and the corresponding magnetisation dynamics. The laser power energy used is 2.5e21$J/m^3s$ and a 50fs pulse duration. 
 }
\label{fig_checkerboard_switching}
\end{figure*}

We begin by constructing a simulation that mimics the experimental scenario, but in a controlled and simplified way. Specifically, the experimental samples, GdFeCo, have regions of different chemical composition (inhomogeneities) that change continuously~\cite{iacocca2019spin,Graves2013}.
A further complication is that high intensity ultrashort laser pulses lead to non-uniform heating effects, particularly when investigating thin films where, due to a (usually) small optical penetration depth of the laser, the surface can be heated to several hundred degrees higher than atomic planes lying tens of nanometres below \cite{Chen2006,jiang2005improved}.
It is known from spin caloritronics that temperature gradients can excite a magnon current from hot to cold regions~\cite{Bauer2012}.
However, on the ultrashort time-scale  (<1ps) it is not clear how this process occurs and what are the time and length-scales involved. Specifically, as the spin, phonon and itinerant electron subsystems may be out-of-equilibrium, how do we quantify a temperature and what is the role of each on these magnon currents? 

To quantify the contribution of magnon currents we use atomistic spin dynamics simulations of laser-induced excitation \cite{evans2014atomistic,Ostler2011}. Our formalism relies on a Heisenberg Hamiltonian and as such does not take into account spin transport effects so that we can disentangle at least part of this complicated picture. Previous simulations have investigated films with a single concentration of Gd with a homogenous distribution of Gd~\cite{Barker2013}. For each given amount of Gd, the minimum laser power required to induce switching is different. In our case, we are considering two different compositions: of 20\% Gd and 30\% Gd. The 30\% sample switches at a lower laser power than the 20\% case. Therefore, we fix the laser power such that if we had a homogenous 30\% sample it would switch, but the fluence would not be sufficient to switch a pure 20\% sample.

We have therefore posed a much simpler structure to investigate the role of inhomogeneities in the ultrafast switching process. Our ``sample'' is constructed with a checkerboard structure, where each square has a different composition.
Now, if we consider the case where each square is large (as in figure \ref{fig_checkerboard_switching}a) the magnetisation dynamics are unaffected by the interface between the 20\% and 30\% Gd regions, they essentially behave as independent regions, on this time scale. On the longer time-scale, we would expect that due to the exchange at the interface, all spins on a given sublattice would re-align due to domain wall motion. As the squares become smaller (shown schematically in inset of Fig.~\ref{fig_checkerboard_switching}, what we observe (Fig.~\ref{fig_checkerboard_switching} left to right), is a change in the behaviour whereby both composition regions reverse on the ultrashort time-scale. 
Our hypothesis is that, during the laser excitation, the two regions will absorb different amounts of energy, leading to an imbalance in the magnon distribution, which can be thought of as an imbalance in energy of the spin system or equivalently the effective spin temperature.
As we do not implicitly simulate the phonons or itinerant electron system, the imbalance leads to rapid transfer of energy via spin angular momentum, producing a net magnon current between the regions on the ultrafast time scale.   

To validate this hypothesis two conditions are required: 1) existence of energy imbalance during the laser excitation and 2) rapid transfer of angular momentum via magnon current due to the energy imbalance. The first condition can be validated by considering an effective spin temperature~\cite{Ma2010}, the energy of the system, or (local) magnetisation. All of them provide information about the absorbed thermal energy from the laser into the spin system.
Here we consider an effective spin temperature \cite{Ma2010} that can be determined by considering the spin and field ensembles, $\{S_i\}$ and $\{H_i\}$, respectively:
\begin{align}
    T_s=\frac{\sum_i [S_i \times H_i ]^2}{2K_bT\sum_i [S_i \cdot H_i ]^2}
    \label{eq_spintemp}
\end{align}
where $k_{\mathrm{B}}$ is Boltzmann's constant and $T$ is the bath temperature, which we determine through solutions to the two-temperature model~\cite{Chen2006}. It is important to mention that the spin temperature is not defined for dynamical systems far from equilibrium such as those involved in the ultrafast magnetisation switching. Here, we do not intend to discuss the absolute value of the temperature, but rather use temperature difference as a relative measure of energy imbalance.  

\begin{figure}[!tb]
\centering
     \includegraphics[angle = -0,width = 1\linewidth]{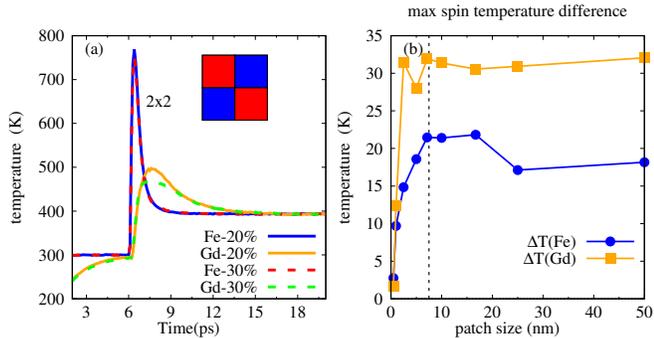}
  \caption[]{
 (a) Example of spin temperature evolution for 2x2 checkerboard under laser pulse. The initial 5ps are shown to indicate the equilibration of the system to room temperature (300K). The spin temperature is computed using eq. \ref{eq_spintemp}. (b) The maximum spin temperature difference for each sub-lattice as a function of square size. }
\label{fig_checkerboard_temperature}
\end{figure}

We have computed the spin temperature for each region and each sub-lattice individually. From the time evolution of the process 
(fig \ref{fig_checkerboard_temperature}a), we extract the maximum temperature difference between the two regions (fig. \ref{fig_checkerboard_temperature}b). This clearly shows that the two regions will absorb a different amounts of energy from the laser pulse and, at the interface, there will be an energy imbalance which will induce a re-equilibration process between the two regions.
In our model, the energy imbalance must be addressed by the transfer of magnons due to the localised nature of the atomic spins.

In panel b of Fig.~\ref{fig_checkerboard_temperature}, we have marked (dashed black line)   the point where switching begins to occur.  Closer inspection of the data shows, not unexpectedly, that the higher effective spin temperature in the 20\% case decreases and the lower effective spin temperature (the 30\% area) increases.
Below this line, the switching occurs as the size of each region is smaller and the difference in spin temperature tends towards zero. This demonstrates a typical length-scale of around 7-8nm (or lower as the propagation is from multiple directions) for the ultrafast magnon propagation due to spin temperature imbalance that can act as a direct energy transfer to drive switching. This is in agreement with the experimental results from ref~\cite{iacocca2019spin} on chemical inhomogeneities in GdFeCo .

The results of Figs.~\ref{fig_checkerboard_switching} and~\ref{fig_checkerboard_temperature} indicate that magnons are being excited and are propagating from one region to the other on the ultrafast time-scale. However, it is not clear from the data presented so far exactly how these magnons propagate. Specifically; how quickly and how far do they propagate; and are they in the diffusive or ballistic regimes, or somewhere in between, i.e. the so-called superdiffusive regime? According the the theory of anomolous diffusion, the variance, $\sigma^2$, of the displacement of a particle (in our case a magnon) is given by \cite{bouchaud1990anomalous,artuso1993periodic,metzler2000random}:
\begin{equation}
    \sigma^2(t) = K_w t^{2/d_m}
\end{equation}
where, $K_m$ represents the generalised diffusion coefficient and $d_m$ is called the Anomolous Diffusion Exponent. Ballistic diffusion is characterised by a constant velocity (no scattering) with a diffusion exponent, $d_m=1$. Standard thermal diffusion processes governed by brownian motion corresponds to a linear growth in $\sigma^2$ and hence has a diffusion exponent equal to 2. Between these two values is the \textit{so-called} superdiffusive regime.

\begin{figure}[!tb]
\centering
     \includegraphics[angle = -0,width = \linewidth]{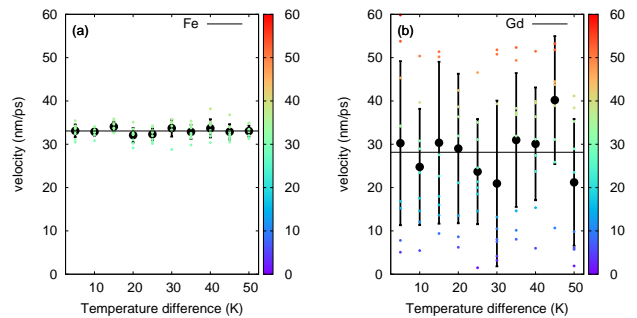}
  \caption[]{
 Magnon propagation velocity for Fe (a) and Gd (b). Due to the thermal fluctuation, there is a distribution of magnons that propagate at the interface and therefore have a distribution of velocities. Averaging over 10 independent calculations is performed to compute the average velocity (illustrated with black dots) and the individual velocities shown on the colour map. Red indicate high velocity and blue corresponds to low velocities.   }
\label{fig_magnon_vel}
\end{figure}

As our model uses fluctuating stochastic fields to mimic thermal fluctuations and, due to the large temperature changes of the thermal bath due to the laser excitation, it is extremely challenging to identify individual magnons or a wavefront associated with the excitation of magnons across the interface, as their amplitude is relatively small compared to the background from thermal noise variation ($\Delta T \approx$1000K  in just 2ps). Therefore, to understand the magnon transport processes we propose the following scenario. We construct a bilayer system with an interface that contains 20\% Gd on one side and 30\% Gd on the other side. We set the temperature of the 30\% layer to 0K and that of the hot, 20\% layer, is set to $T_H$. They are initially decoupled to allow for thermal equilibration for a period of 1ps to ensure that the spin temperatures are equal to that of their respective thermal baths, after which we couple the systems to their respective thermal baths (at 0K and $T_H$, respectively) and monitor the resulting spin dynamics. We use temperature difference up to around the values found for the spin temperature differences in Fig.~\ref{fig_checkerboard_temperature}.

\begin{figure*}[!htb]
\centering
     \includegraphics[angle = -0,width = 1\linewidth]{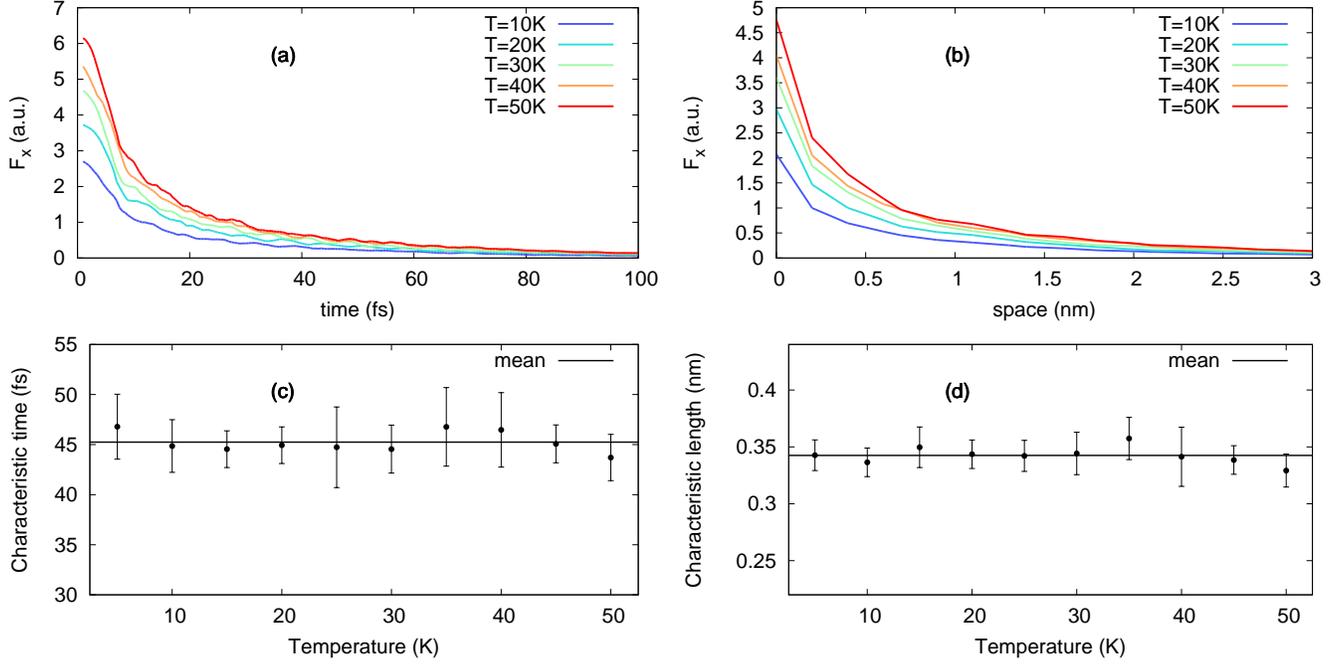}
  \caption[]{
Magnon flux evolution in time (a) and in space (b). (a) and (b) illustrate the propagation of the maximum of the magnon flux. The characteristic length (c) and time (d) for the magnon flux is computed as integral propagation length/relaxation time.}
\label{fig_magnon_flux}
\end{figure*}

The two regions will have magnons with different energies, leading to an imbalance.  To re-equilibrate there will be a net magnon (energy) transfer mediated by the exchange interaction as was seen in the checkerboard. 
Thus, high energy magnons from the hot regions will cross the interface and propagate into the cold region. The propagation can be visualised via the change in the net magnetisation ($\Delta m$) with respect to thermal equilibrium value. 
To calculate the magnon transport through the interface, and the propagation of magnons, we used the expression for the energy flux vector similar to the Poynting vector in electromagnetism:
\begin{align}
 \bold{F} = \frac{d (\Delta \bold{m} )}{dt} \times \Delta \bold{m} ~,~~
 \Delta \bold{m}=\sum | \bold{s}-\bold{m_{eq}} |  
\end{align}
where $\Delta \bold{m}$ is the net magnetisation as a function of $x$, the direction perpendicular to the interface and  $d\bold{m}$(x)/dt is the rate of change of magnetisation in time. The approach is similar to the one used in spin pumping~\cite{Bauer2012}. $\mathbf{F}$ gives the net flux through a given plane (the interface plane, YZ plane, in our case).  The maximum change in $\mathbf{F}$ is correlated with the front of magnon propagation.

By calculating the flux vector we can track the propagating wavefront as magnons are moving from one region to another.
Analysis of the data shows that the wavefront initially propagates linearly, meaning that the magnons on the ultrafast time-scale are propagating ballistically. The velocity of the magnons is straightforwardly extracted from a linear fit to the data and is independent of the temperature difference (Fig.~\ref{fig_magnon_vel}).
It is important to note that the ballistic nature of the magnon transport at the interface is on the femtosecond time scale, with velocities around 30nm/ps. Such large velocities indicate that on the time scale relevant for ultrafast switching (<1ps), the magnon current can have a significant role in the switching of composite systems and multi-layer structures. 
At the same time, the magnons do not, however, propagate with a constant amplitude as energy is transferred from the thermally excited spins to those with low thermal energy.

To investigate the amplitude of the magnon propagation we extract the time (fig. \ref{fig_magnon_flux}a) and spatial (fig. \ref{fig_magnon_flux}b) evolution of the front/maximum of the flux.
It is important to highlight that this is an average property over the entire interface. 
Due to the thermal origin of the magnons, there will be a distribution of magnons crossing the interface, and the amplitude of the energy transfer will be proportional to the temperature (thermal energy) in the hot region.
We compute the integral characteristic time \cite{Garanin1999}
as $\int_0^{\infty} \frac{F_x(t)}{F_x(t=0)}dt $, where the normalization to the initial value ($F_x(t=0)$) decouples the characteristic time from the initial conditions. A similar consideration is done for magnon propagation in space. Both the characteristic time and length scales are independent of the temperature difference between regions (fig.~\ref{fig_magnon_flux}) and therefore are related to material properties. 
The results in fig.~\ref{fig_magnon_flux} show, that ultrafast magnons can be generated due to spin a temperature imbalance and the propagation timescale is fast enough to play an important role in the switching process at the 100fs-ps time scale. The characteristic time scale of around 45fs is similar to the case of charge driven spin currents reported in Ref.~\cite{battiato2012theory}. These results are of significant importance indicating that both charge and magnon driven angular moment are playing a role in the ultrafast switching of magnetic systems. The advantage of magnon driven mechanism is that it manifests in a similar strength between metal and insulator materials such as YIG, whereas charge driven mechanism is restricted to metallic materials. 

The results indicate that using chemical inhomogeneities can expand the rare-earth concentration range for deterministic all-optical switching by using the interface magnon currents to compensate the energy required for bridging the FM-AFM states. This can explain the experimental results in GdFeCo thin film samples, where such inhomogeneities have been investigated~\cite{Graves2013, iacocca2019spin}.
The very localised nature of the mechanism is promising for utilization in nanoscale devices. 
This could lead to a new design of systems that undergo all-optical switching. In particular, combining structures with different functionalities will lead to improved overall performance. One example would be to combine low-laser power switching materials, such as GdFeCo, 
with high anisotropy composites such as FePt, which does not undergo AOS, but provides thermal stability, and can provide a solution for memory applications based on all-optical switching. 
From a practical point of view, currently, checkerboard structures may be experimentally hard to achieve. An alternative could consist of engineering thin bilayer and multilayer structures with thicknesses on the length scale of the magnon propagation, which are easier to fabricate. The time-scale is governed by the strength of the exchange interaction, which could be controlled by the choice of interface material.

\begin{acknowledgments}
Thomas Ostler gratefully acknowledges the Vice Chancellor's Fellowship Scheme at Sheffield Hallam University for supporting this research. 
The work has been facilitated by the COST Action CA17123 Magnetofon.
This project has received funding from the European Unions Horizon 2020 research and innovation programme under Grant Agreement No. 737093 (FEMTOTERABYTE). The atomistic simulations were undertaken on the VIKING cluster, which is a high performance compute facility provided by the University of York.
\end{acknowledgments}

\end{document}